\documentclass[twoside,leqno,twocolumn]{article}  
\usepackage{ltexpprt} 
\usepackage{amsfonts}
\usepackage{amsmath,amssymb}
\usepackage{hyperref}

\begin{document}

\title{\Large The Hunting of the Bump: On Maximizing Statistical Discrepancy}
\author{Deepak Agarwal\thanks{AT\&T Labs -- Research \texttt{dagarwal@research.att.com}} \\
\and 
Jeff M. Phillips\thanks{Duke University \texttt{jeffp@cs.duke.edu}, supported by a James B. Duke Graduate Fellowship and a NSF Graduate Research Fellowship.} \and
Suresh Venkatasubramanian\thanks{AT\&T Labs -- Research \texttt{suresh@research.att.com}}}
\date{}

\maketitle

 \begin{abstract} \small\baselineskip=9pt 
Anomaly detection has important applications in biosurveilance and
environmental monitoring.  When comparing measured data to data drawn from
a baseline distribution, merely, finding clusters in the measured data may
not actually represent true anomalies.  These clusters may likely be the
clusters of the baseline distribution.  Hence, a discrepancy function is
often used to examine how different measured data is to baseline data
within a region.  An anomalous region is thus defined to be one with high
discrepancy.

In this paper, we present algorithms for maximizing statistical
discrepancy functions over the space of axis-parallel rectangles.  We give
provable approximation guarantees, both additive and relative, and our
methods apply to any convex discrepancy function.  Our algorithms work by
connecting statistical discrepancy to combinatorial discrepancy; roughly
speaking, we show that in order to maximize a convex discrepancy function
over a class of shapes, one needs only maximize a linear discrepancy
function over the same set of shapes.

We derive general discrepancy functions for data generated from a one-
parameter exponential family. This generalizes the widely-used Kulldorff
scan statistic for data from a Poisson distribution.  We present an
algorithm running in $O(\smash[tb]{\frac{1}{\epsilon} n^2 \log^2 n})$ that computes the
maximum discrepancy rectangle to within additive error $\epsilon$, for the
Kulldorff scan statistic.  Similar results hold for relative error and for
discrepancy functions for data coming from Gaussian, Bernoulli, and gamma
distributions.  Prior to our work, the best known algorithms were exact
and ran in time $\smash[t]{O(n^4)}$.
\end{abstract}

\section{Introduction}
\label{sec:intro}
Outlier detection (or ``bump hunting''\cite{FF99}) is a common
problem in data mining. Unlike in robust clustering settings, where
the goal is to detect outliers in order to remove them, outliers are
viewed as \emph{anomalous events} to be studied further. In the area of 
biosurveillance for example, an outlier would
consist of an area that had an unusually high disease rate (disease
occurence per unit population) of a particular ailment. In
environmental monitoring scenarios, one might monitor the rainfall
over an area and wish to determine whether any region had unusually
high rainfall in a year, or over the past few years.

A formal statistical treatment of these problems allows us to abstract
them into a common framework. Assume that data (disease rates, rainfall
measurements, temperature) is generated by some stochastic spatial 
process. Points in space are either fixed or assumed to be
generated from some spatial point process and measurements on points are
assumed to be statistically independent and follow a distribution from a
one-parameter exponential family. Also, let $b(\cdot)$ be some baseline
measure defined on the plane. For instance, $b(\cdot)$ can be the
counting measure (counts the number of points in a region), volume measure
(measures volume of a region), weighted counting measure (adds up
non-negative weights attached to points in a region). In biosurveillance,
the counting measure (gives the region population) is often used to
discover areas with elevated disease risk. Weighted counting measures which
aggregate weights assigned to points based on some attribute (e.g. race of
an individual) have also been used (see \cite{kulldorff:comm} for an
example).  Let $p$ be a set of points generating a set of measurements
$m(p)$. Given a measure of discrepancy $f$ that takes as input the
functions $m(\cdot)$ and $b(\cdot)$ and a collection of regions ${\cal
  S}$, the problem of \emph{statistical discrepancy} can be defined as:

\begin{center}
  Find the region $S \in {\cal S}$ for which $f$ invoked on the
  measurements for points in $S$ is maximized.
\end{center}

Statistical discrepancy functions arise by asking the following question:
``How likely is it that the observed points in $S$ come from a
distribution that is different than the distribution of points in
$S^{c}$?''. The function $f$ is derived using a
\emph{likelihood ratio test} which has high statistical power to detect
the actual location of clusters, but is computationally difficult to deal
with.  As a consequence, most algorithmic work on this problem has
focused either on fast heuristics that do not search the entire space of
shapes, or on conservative heuristics that guarantee finding the maximum
discrepancy region and will often (though not always) run in time less
than the worst-case bound of $|{\cal S}|$ times the function evaluation
cost.

Apart from identifying the region for which $f$ is maximized, it is
customary to evaluate the likelihood of the identified cluster being
generated by chance, i.e., compute the probability (called p-value) of
maximum discrepancy to exceed the observed maximum discrepancy under the
null hypothesis of no clustering effect.  A small p-value (e.g. $< .05$)
will mean the identified cluster is statistically significant.  Since the
distribution of $f$ is usually not analytically tractable, randomization tests
(\cite{dwass,good:book}) which involve multiple instances of the maximum
discrepancy computation are run for data generated from the null model.
Thus, the computation of statistical discrepancy is the main algorithmic
bottleneck and is the problem we focus on in this paper.
%Section~\ref{sec:concl} discusses the potential value of solving decision
%variants of the problem (``Is discrepancy $\le k$ ?'').

\subsection{Our Contributions}
\label{ssec:contrib}

In this paper, we present algorithms with non-trivial worst-case running
time bounds for approximating a variety of statistical discrepancy
functions. Our main result is a structural theorem that reduces the
problem of maximizing any convex discrepancy function over a class of
shapes to maximizing a simple linear discrepancy function over the same
class of shapes. 

The power of this theorem comes from the fact that there are known
algorithms for maximizing special kinds of linear discrepancy functions,
when the class of shapes consists of axis-parallel rectangles. Given two
sets of red and blue points in the plane, the \emph{combinatorial
  discrepancy} of a region is the absolute difference between the number
of red and blue points in it. Combinatorial discrepancy is very valuable
when derandomizing geometric algorithms; it also appears in machine
learning as the relevant function for the \emph{minimum disagreement
  problem}, where red and blue points are thought of as good and bad
examples for a classifier, and the regions are hypotheses.  This problem
(and a more general variant of it) was considered by Dobkin, Maass and
Gunopoulos in 1995~\cite{DGM95}, where they showed that combinatorial
discrepancy for axis-parallel rectangles in the plane could be maximized
exactly in time $O(n^2\log n)$, far better than the $O(n^4)$ bound that a
brute-force search would entail.

We show that the Dobkin~\emph{et. al.} algorithm can be extended fairly easily to
work with general linear discrepancy functions. This result, combined with
our general theorem, allows us to approximate \emph{any} convex
discrepancy function over the class of axis-parallel rectangles. We
summarize our results in Table~\ref{tab:results}; as an
example, we present an additive approximation algorithm for the Kulldorff
scan statistic that runs in time $O(\frac{1}{\epsilon}n^2\log^2 n)$, which
compares favorably to the (trivial) $O(n^4)$ running time to compute an
exact solution.

Essentially, the reduction we use allows us to decouple the measure of
discrepancy (which can be complex) from the shape class it is maximized
over. Using our approach, if you wanted to maximize a general discrepancy
function over a general shape class, you need only consider combinatorial
discrepancy over this class. As a demonstration of the generality of our
method, we also present algorithms for approximately maximizing
discrepancy measures that derive from different underlying distributions.
In fact, we provide general expressions for the one-parameter exponential
family of distributions which includes Poisson, Bernoulli, Gaussian and
Gamma distributions. For the Gaussian distribution, the measure of
discrepancy we use is novel, to the best of our knowledge. It is derived
from maximum likelihood considerations, has a natural interpretation as a
$\chi^2$ distance, and may be of independent interest.

Another notion of outlier detection incorporates a time dimension. In
\emph{prospective outlier detection}, we would like to detect the maximum
discrepancy region over all time intervals starting from the present and
going backwards in time. We show that linear discrepancy can be maximized
over such time intervals and, as a consequence of our reduction, show that
any convex discrepancy function can be approximately maximized. 

\begin{table*}[thbp]
  \centering
  \begin{tabular}{|c|c|c|c|} \hline
    & \multicolumn{2}{|c|}{This paper} & Prior work \\ \hline
    & OPT $-\epsilon$ & $\textrm{OPT}/(1+\epsilon)$ & Exact \\ \hline
    Poisson (Kulldorff)/Bernoulli/Gamma & $O(\frac{1}{\epsilon}n^2\log^2 n)$ & $O(\frac{1}{\epsilon}n^2\log^2 n)$ & $O(n^4)$ \\ \hline
    Gaussian & $O(\frac{1}{\epsilon}n^3\log n\log \log n)$ &
    $O(\frac{1}{\epsilon}n^2\log^2 n)$ & $O(n^4)$ \\ \hline
  \end{tabular}
  \caption{Our results. For prospective discrepancy, multiply all bounds
    by $n$, and for higher dimensions, multiply by $n^{2d-4}$.}
  \label{tab:results}
\end{table*}

\section{Related Work}
\label{sec:related}

Detecting clustering effects in spatial data is a well-studied problem in
statistics\footnote{It goes without saying that there is a huge literature
  on clustering spatial data. Since our focus is primarily on
  statistically sound measures, a survey of general clustering methods is
  beyond the scope of this paper.}. Much of the early focus has been on
devising efficient statistical tests to detect presence of clustering at a
global level without emphasis on identifying the actual clusters (see
\cite[Chapter 8]{cressie}).  The spatial scan statistic, introduced by
Kulldorff~\cite{kulldorff:comm} provides an elegant solution for detection
and evaluation of spatial clusters.  The technique has found wide
applicability in areas like public health, biosurveillance, environmental
monitoring \emph{etc}. For interesting applications and detailed
description of scan statistics, we refer the reader to
\cite{glazbala,glaz}.  Generalization of the spatial scan statistic to a
space-time scan statistic for the purpose of prospective surveillance has
been proposed by Kulldorff~\cite{kulldorffprospective}, and Iyengar~\cite{iyengar}
suggested the use of ``expanding-in-time'' regions to detect space-time
clusters. We note that the algorithms described by Kulldorff are
heuristics: they do not guarantee any bound on the quality of the
solution, and do not traverse the entire space of regions. The regions he
considers are circular, and cylindrical in the case of prospective
surveillance.

Dobkin and Eppstein~\cite{DE93} were the first to study efficient
algorithms to compute maximum discrepancy over a range space. Their
algorithms compute discrepancy in a region $R$ as a difference between the
fraction of points in $R$ and the fraction of the total area represented
by $R$.  This measure stems from evaluating fundamental operations for
computer graphics. Their ranges were half spaces and axis-oriented
orthants centered at the origin, limited to the unit cube, and their
results extended to $d$-dimensional spaces. Subsequently Dobkin,
Gunopulous, and Maass \cite{DGM95} developed algorithms for computing
maximum bichromatic discrepancy over axis-parallel rectangular regions,
where the bichromatic discrepancy of a region is the difference between
the number of red points and the number of blue points in the region.
This solves the \emph{minimum disagreement problem} from machine learning,
where an algorithm finds the region with the most \emph{good} points and
the fewest \emph{bad} points, a key subroutine in agnostic PAC learning. 

Recently, Neill and Moore have developed a series of algorithms to
maximize discrepancy for measures such as the Kulldorff spatial scan
statistic. Their approach works for axis parallel squares~\cite{NM04b} and
rectangles~\cite{NM04}.  Their 
algorithms are conservative, in that they always find the region of
maximum discrepancy. The worst-case running time of their algorithms is
$O(n^4)$ for rectangles and $O(n^2)$ for fixed-size squares since the
algorithms enumerate over all valid regions. However, they use efficient
pruning heuristics that allow for significant speedup over the worst case
on most data sets. An alternative approach by Friedman and Fisher
\cite{FF99} greedily computes a high discrepancy rectangle, but has no
guarantees as to how it compares to the optimal. Their approach is quite
general, and works in arbitrary dimensional spaces, but is not
conservative: many regions will remain unexplored.

A one-dimensional version of this problem has been studied in
bioinformatics~\cite{LABLY05}.  The range space is now the set of all
intervals, a problem with much simpler
geometry.  In this setting, a relative $\epsilon$-approximation can be
found in $O(\frac{1}{\epsilon^2} n)$ time.

A related problem that has a similar flavor is the so-called \emph{Heavy
  Hitters} problem~\cite{CM03,CKMS04}. In this problem, one is given a
multiset of elements from a universe, and the goal is to find elements
whose frequencies in the multiset are unusually high (i.e much more than
the average). In a certain sense, the heavy hitter problem fits in our
framework if we think of the baseline data as the uniform distribution,
and the counts as the measurements. However, there is no notion of
ranges\footnote{Hierarchical heavy hitters provide the beginnings
  of such a notion.} and the heavy hitter problem itself is interesting in
a streaming setting, where memory is limited; if linear memory is
permitted, the problem is trivial to solve, in contrast to the problems we
consider.

\section{Preliminaries}
\label{sec:preliminaries}

Let $P$ be a set of $n$ points in the plane. Measurements and baseline
measures over $P$ will be represented by two functions, $m : P
\rightarrow \mathbb{R}$ and $b : P \rightarrow \mathbb{R}$. ${\cal R}$ denotes a
range space over $P$. A \emph{discrepancy function} is defined as $d: (m,
b, R) \rightarrow \mathbb{R}$, for $R \in {\cal R}$.

%The discrepancy function is defined generally in this way to capture
%different kinds of clustering. Note that if $m,b : P \rightarrow \{0,1\}$
%merely label points of $P$ as red or blue, and if we define $ d_C(m, b, R)
%= |\sum_{p \in R} m(p) - b(p)| $ we obtain the standard definition of
%combinatorial (bichromatic) discrepancy, where the discrepancy of a region
%is the difference between the number of red and blue points inside it.
Let
$m_R = \sum_{p \in R} m(p)/M, b_R = \sum_{p \in R} b(p)/B$, where $M =
\sum_{p \in U} m(p)$, $B = \sum_{p \in U} b(p)$, and $U$ is some box
enclosing all of $P$. %We can rewrite $d_C(m,p,R) = | M * m_R - B * b_R|$.
\emph{We will assume that $d$ can be written as a convex function of $m_R,
  b_R$}. All the discrepancy functions that we consider in this paper
satisfy this condition; most discrepancy functions considered
prior to this work are convex as well. We can write $d(m, b, R)$
as a function $d' : [0,1]^2 \rightarrow \mathbb{R}$, where $d(m, b, R) =
d'(m_R, b_R)$. We will use $d$ to refer to either function where the
context is clear.

\emph{Linear discrepancy functions} are a special class of discrepancy
functions where $d = \alpha \cdot m_R + \beta \cdot b_R + \gamma$. It is easy to see that
combinatorial (bichromatic) discrepancy, the difference between the number
of red and blue points in a region, is a special case of a linear
discrepancy function. 

The main problem we study in this paper is:
\begin{problem}[Maximizing Discrepancy]
  Given a point set $P$ with measurements $m$, baseline measure $b$, a  range space ${\cal R}$, and a
  convex discrepancy function $d$, find the range $R \in {\cal R}$ that
  maximizes $d$. 
\end{problem}

An equivalent formulation, replacing the range $R$ by the point $r = (m_R,
b_R)$ is:
\begin{problem}
  Maximize convex discrepancy function $d$ over all points $r = (m_R,
  b_R), R \in {\cal R}$. 
\end{problem}

Assume that points now arrive with a timestamp $t(\cdot)$, along with the
measurement $m(\cdot)$ and baseline $b(\cdot)$. In \emph{prospective
  discrepancy} problems, the goal is to maximize discrepancy over a region
in space and time defined as $R \times [t, t_{\text{now}}]$, where $R$ is
a spatial range. In other words, the region includes all points with a
timestamp between the present time and some time $t$ in the past. Such
regions are interesting when attempting to detect \emph{recent} anomalous
events.

\begin{problem}[Prospective discrepancy]
  Given a point set $P$ with measurements $m$, baseline measure $b$,
  timestamps $t$, a  range space ${\cal R}$, and a
  convex discrepancy function $d$, find the range $T = (R, [t^*,\infty]), R \in {\cal R}$ 
that maximizes $d$.   
\end{problem}

\subsection{Boundary Conditions}
\label{ssec:boundary-conditions}

As we shall see in later sections, the discrepancy functions we consider
are expressed as log-likelihood ratios. As a consequence, they tend to
$\infty$ when either argument tends to zero (while the other remains
fixed). Another way of looking at this issue is that regions with very low
support often correspond to overfitting and thus are not interesting.
Therefore, this problem is typically addressed by requiring a
\emph{minimum level of support} in each argument.  Specifically, we will
only consider maximizing discrepancy over shapes $R$ such that $m_R > C/n,
b_R > C/n$, for some constant $C$. In mapping shapes to points as described
above, this means that the maximization is restricted to points in the
square $S_n = [C/n,1-C/n]^2$.  For technical reasons, we will also assume
that for all $p$, $m(p), b(p) = \Theta(1)$. Intuitively this reflects the
fact that measurement values are independent of the number of observations
made.

\section{A Convex Approximation Theorem}
\label{sec:conv-appr-theor}

We start with a general approximation theorem for maximizing a convex
discrepancy function $d$. Let $\ell(x,y) = a_1x + a_2y + a_3 $ denote a
linear function in $x$ and $y$. Define an \emph{$\epsilon$-approximate
  family} of $d$ to be a collection of linear functions $\ell_1, \ell_2,
\ldots, \ell_t$ such that $l^U(x,y) = \max_{i \le t} \ell_i(x,y)$, the
\emph{upper   envelope} of the $\ell_i$, has the property that    
 $l^U(x,y) \le d(x,y) \le l^U(x,y) + \epsilon $
Define a \emph{relative} $\epsilon$-approximate family of $d$ to be a
collection of linear functions $\ell_1, \ell_2, \ldots, \ell_t$ such that
$ l^U(x,y) \le d(x,y) \le (1+\epsilon)l^U(x,y) $

\begin{lemma}
\label{lemma:convex2linear}
Let $\ell_1, \ell_2, \ldots, \ell_t$ be an $\epsilon$-approximate family of a convex discrepancy function $d: [0,1]^2 \rightarrow \mathbb{R}$.
Consider any point set $\mathcal{C} \subset [0,1]^2$.
Let $(x^*_i,y^*_i) = \arg\max_{\mathbf{p} \in \mathcal{C}} \ell_i(\mathbf{p}_x,\mathbf{p}_y)$, 
and let $(x^*,y^*) = \arg  \max_{x^*_i,y^*_i} \ell_i(x^*_i, y^*_i)$. 
Let $d^* = \sup_{\mathbf{p} \in \mathcal{C}} d(\mathbf{p}_x,\mathbf{p}_y)$,
$d_{\inf} = \inf_{\mathbf{q} \in [0,1]^2} d(\mathbf{q}_x,\mathbf{q}_y)$ and
let $m = \max(l^U(x^*,y^*),d_{\inf})$ . 
Then \[ m \le d^* \le m + \epsilon \]

If $\ell_1, \ell_2, \ldots, \ell_t$ is a relative $\epsilon$-approximate family, then
  \[ m \le d^* \le (1+\epsilon)m  \]
\end{lemma}
\begin{proofsketch}
By construction, each point $(x^*_i,y^*_i,l_i(x^*_i, y^*_i))$ lies on the
upper envelope $l^U$. The upper envelope is convex, and lower bounds
$d(\cdot)$, and therefore in each \emph{patch} of $l^U$ corresponding to a
particular $\ell_i$, the maximizing point $(x^*_i,y^*_i)$ also maximizes
$d(x,y)$ in the corresponding patch. This is only false for  the
patch of $l^U$ that supports the minimum of $d(x,y)$, where 
the term involving $d_{\inf}$ is needed. This corresponds to adding a
single extra plane tangent to $d(\cdot)$ at its minimumm, which is unique
by virtue of $d(\cdot)$ being convex.
\end{proofsketch}

\begin{lemma}
\label{lemma:hessian1}
Let $f : [0,1]^2 \rightarrow \mathbb{R}$ be a convex smooth function. Let
$\tilde{f} : [0,1]^2 \rightarrow \mathbb{R}$ be the linear approximation to
$f$ represented by the hyperplane tangent to $f$ at $\mathbf{p} \in
[0,1]^2$.  Then $\tilde{f}(\mathbf{p}) \le f(\mathbf{p})$, and $f(\mathbf{p}) -
\tilde{f}(\mathbf{q}) \le \| \mathbf{p} -\mathbf{q} \|^2\lambda^*$, where
$\lambda^*$ is the maximum value of the largest eigenvalue of $H(f)$,
maximized along the line joining $\mathbf{p}$ and $\mathbf{q}$.
\end{lemma}

\begin{proof}
  $\tilde{f}(\mathbf{q}) = f(\mathbf{p}) + (\mathbf{q} - \mathbf{p})^\top
  \nabla f(\mathbf{p})$. The inequality $\tilde{f}(\mathbf{p}) \le
  f(\mathbf{p})$ follows from the convexity of $f$. By Taylor's theorem
  for multivariate functions, the error $f(\mathbf{p}) - \tilde{f}(\mathbf{q}) =
  (\mathbf{q-p})^\top H(f)(\mathbf{p}^*) (\mathbf{q-p})$, where $H(f)$ is the
  Hessian of $f$, and $\mathbf{p}^*$ is some point on the line joining  $\mathbf{p}$ and $\mathbf{q}$.

  Writing $\mathbf{q-p}$ as $\|\mathbf{q} - \mathbf{p}\|\mathbf{\hat{x}}$,
  where   $\mathbf{\hat{x}}$ is a unit vector, we see that the error is
  maximized when the expression $\mathbf{\hat{x}}^\top H(f)
  \mathbf{\hat{x}}$ is maximized, which happens when $\mathbf{\hat{x}}$ is
  the eigenvector corresponding to the largest eigenvalue $\lambda^*$ of
  $H(f)$.  
\end{proof}

Let $\lambda^* = \sup_{\mathbf{p} \in S_n} \lambda_{\max}(H(f)(\mathbf{p}))$. Let $\epsilon_\mathbf{p}(\mathbf{q}) = \| \mathbf{p} -\mathbf{q}
\|^2\lambda^*, \epsilon^R_\mathbf{p}(\mathbf{q}) = \| \mathbf{p} -\mathbf{q}
\|^2\lambda^*f(\mathbf{p})$. 

\begin{lemma}
\label{lemma:planes2points}
  Let ${\cal C} \subset S_n$ be a set of $t$ points such that for all
  $\mathbf{q} 
  \in S_n, \min_{\mathbf{p} \in {\cal C}}
  \epsilon_\mathbf{p}(\mathbf{q}) 
  (\textrm{resp. } \epsilon^R_\mathbf{p}(\mathbf{q})) \le
  \epsilon$. Then the $t$ tangent planes at the points $f(\mathbf{p}),
  \mathbf{p} \in {\cal C}$ form an $\epsilon$-approximate (resp. relative
  $\epsilon$-approximate) family for $f$. 
\end{lemma}
\begin{proof}
  Let ${\cal C} = \{\mathbf{p}_1, \ldots, \mathbf{p}_t\}$. Let $l_i$ denote the
  tangent plane at $f(\mathbf{p}_i)$. For all $i$, $l_i(\mathbf{q}) \le
  f(\mathbf{q})$ by Lemma~\ref{lemma:hessian1}. Let $j = \arg\min_i
  \epsilon_\mathbf{p_i}(\mathbf{q})$. Then $l_j(\mathbf{q}) = \max_i
  l_i(\mathbf{q}) \le \epsilon$. A similar argument goes through for
  $\epsilon^R_\mathbf{p_i}(\mathbf{q})$ 
\end{proof}

The above lemmas indicate that in order to construct an
$\epsilon$-approximate family for the function $f$, we need to sample an
appropriate set of points from $S_n$. A simple area-packing bound,
using the result from Lemma~\ref{lemma:planes2points}, indicates that we would
need $O(\lambda^*/\epsilon)$ points (and thus that many planes). 
However, $\lambda^*$ is a function of $n$. A stratified grid decomposition
can exploit this dependence to obtain an improved bound. 

\begin{theorem}
  \label{thm:main-approx}
Let $f : [0,1]^2 \rightarrow \mathbb{R}$ be a convex smooth function, and fix
$\epsilon > 0$. Let $\lambda(n) = \lambda^*(S_n)$. Let $F(n, \epsilon)$ be
the size of an $\epsilon$-approximate family for $f$,  and let $F^R(n,
\epsilon)$ denote the size of a relative $\epsilon$-approximate
family. Let $\lambda(n) = O(n^d)$. Then,  
\[ F(n, \epsilon) =   
\begin{cases} 
  O(1/\epsilon) & d = 0 \\
  O(\frac{1}{\epsilon} \log_{\frac{1}{d}}\log n) & 0 < d < 1\\
  O(\frac{1}{\epsilon}\log n) & d = 1 \\
  O(\frac{1}{\epsilon}n^{d-1}\log_d \log n) & d > 1 
\end{cases}
\] 

Let $\lambda'(n) = \lambda(n)/f_{\max}(n)$, where $f_{\max}(n)$ denotes
$\max_{\mathbf{p} \in S_n} f(\mathbf{p})$. Then $F^R(n, \epsilon)$ has
size chosen from the above cases according to $\lambda'(n)$.
\end{theorem}
\begin{proof}
  The relation between $F^R(n,\epsilon)$ and $F(n,\epsilon)$ follows
  trivially from the relationship between $\epsilon^R_\mathbf{p}(\mathbf{q})$
  and $\epsilon_\mathbf{p}(\mathbf{q})$. 
  
  If $\lambda(n)$ is $O(1)$ , then $\lambda^*$ can be upper bounded by a
  constant, resulting in an $\epsilon$-approximate family of size
  $O(1/\epsilon)$.  The more challenging case is when $\lambda^*$ is an
  increasing function of $n$.
  
  Suppose $\lambda^* = O(n^d)$ in the region $S_n$. Consider the following
  adaptive gridding strategy. Fix numbers $n_0, n_1, \ldots n_k$, $n_{k+1}
  = n$. Let $A_0 = S_{n_0} = [1/n_0, 1-1/n_0]^2$. Let $A_i = S_{n_i} -
  \cup_{j < i} A_i$. Thus, $A_0$ is a square of side $1-2/n_0$, and each
  $A_i$ is an annulus lying between $S_{n_{i+1}}$ and $S_{n_i}$. $A_0$ has
  area $O(1)$ and each $A_i, i > 0$ has area $O(1/n_{i-1})$. In each region
  $A_i$, $\lambda^*(A_i) = O(n_i^d)$.
  
  How many points do we need to allocate to $A_0$? A simple area bound
  based on Lemma~\ref{lemma:planes2points} shows that we need
  $\lambda^*(A_0)/\epsilon$ points, which is $O(n_0^d/\epsilon)$.  In each
  region $A_i$, a similar area bound yields a value of $O(n_i^d/\epsilon
  n_{i-1})$. Thus the total number of points needed to construct the
  $\epsilon$-approximate family is $ N(d,k) = n_0^d/\epsilon + \sum_{0 < i \le k+1} n_i^d/\epsilon n_{i-1}$.

Balancing this expression by setting all terms equal, and setting $l_i =
\log n_i$, we obtain the recurrence

\begin{eqnarray}
  l_i &=& \frac{(d+1) l_{i-1} - l_{i-2}}{d}   \label{eq:recurrence1}\\
  l_1 &=& \frac{d+1}{d} l_0 \label{eq:recurrence2}
\end{eqnarray}

\begin{claim}
\label{claim:1}
$  l_{k+1} = \log n = ( 1 + \sum_{i=1}^j d^{-i}) l_{k-j+1} -
(\sum_{i=1}^j d^{-i})l_{k-j}$
\end{claim}
\begin{proof}
%By induction. We omit the details.
  The proof is by induction. The statement is true for $j = 1$ from
  Eq.(\ref{eq:recurrence1}). Assume it is true upto $j$. Then
\begin{eqnarray*}
\nonumber
l_{k+1} &=&
     \left( 1 + \sum_{i=1}^j d^{-i}\right) l_{k-j+1} - 
     \left(\sum_{i=1}^j  d^{-i}\right)l_{k-j} 
\\ &=&   
     \left( 1 + \sum_{i=1}^j d^{-i}\right)\left[ \frac{(d+1) l_{k-j} - l_{k-j-1}}{d} \right] - 
\\ &&
     \left(\sum_{i=1}^j d^{-i}\right) l_{k-j}  
\\ &=&  
     \left( 1 + \sum_{i=1}^{j+1} d^{-i}\right) l_{k-j} - 
     \left(\sum_{i=1}^{j+1} d^{-i}\right)  l_{k-j-1}
\end{eqnarray*}
\end{proof}

Setting $j = k$ in Claim~\ref{claim:1} yields the expression
$ \log n = (1 + \sum_{i=1}^k d^{-i}) l_1 - (\sum_{i=1}^k d^{-i}) l_0 $.
Substituting in the value of $l_1$ from Eq.(\ref{eq:recurrence2}),
$  \log n = (1 + \sum_{i=1}^{k+1} d^{-i}) \log n_0 = 1/\alpha \log n_0 $.
The number of points needed is $F(n,\epsilon) = \frac{k+2}{\epsilon} n_0^d =
\frac{k+2}{\epsilon} n^{d\alpha }$.

How large is $d\alpha$? Consider the case when $d > 1$:
\begin{eqnarray*}
  \frac{d}{(1 + \sum_{i=1}^{k+1} d^{-i})} &= \frac{d-1}{1 - 1/d^{k+2}} = \frac{d^{k+2}}{d^{k+2}-1} (d-1)
\end{eqnarray*}

Setting $k = \Omega(\log_d \log n)$, $F(n,\epsilon) =
O(\frac{1}{\epsilon}n^{d-1}\log_d \log n)$.
For example, $F(n,\epsilon) = O(\frac{1}{\epsilon}n\log\log n)$ when $d = 2$.
Similarly, setting $k = \Omega(\log_{1/d}\log n)$ when $0 < d < 1$
yields $F(n,\epsilon) = O(\frac{1}{\epsilon} \log_{1/d} \log n)$.

When $d = 1$,
$  \frac{d}{(1 + \sum_{i=1}^{k+1} d^{-i})} = \frac{1}{k+2} $. 
Setting $k = \Omega(\log n)$, we get $F(n, \epsilon) = O(\frac{1}{\epsilon}\log n)$.

\end{proof}

\section{Algorithms for Combinatorial Discrepancy}
\label{sec:algos}

\begin{lemma}[\cite{DGM95}] Combinatorial discrepancy for a set of red and blue points in the plane can be computed in time $O(n^2 \log n)$.  
\end{lemma}
\begin{proofsketch}[See~\cite{DGM95} for details]
  A discrepancy maximizing rectangle has minimal and maximal points in the
  x and y dimensions.  These four points fully describe the rectangle.  By
  specifying the minimizing and maximizing y coordinates, the problem is
  reduced to a one-dimensional problem of all points within the slab this
  defines.  By maintaining the maximal discrepancy interval in the
  one-dimensional problem under point insertion, only $O(n^2)$ insertions
  are necessary to check the maximum discrepancy interval over all
  possible slabs, and thus over all possible rectangles.
 
  The one-dimensional problem is solved by building a binary tree of
  intervals.  A node corresponding to interval $I$ stores the
  subinterval $i$ of maximal discrepancy, the interval $l$ of maximal discrepancy that
  includes the left boundary of $I$, and the interval $r$ of maximal discrepancy
  that includes the right boundary of $I$.  Two adjacent nodes, $I_{left}
  : (i_{left}, l_{left}, r_{left})$ and $I_{right} : (i_{right},
  l_{right}, r_{right})$, can be merged to form a single nodes $I :
  (i,l,r)$ in constant time.  $i$ is assigned the interval with the
  maximum discrepancy out of the set $\{i_{left}, i_{right}, r_{left}
  \cup r_{left}\}$.  $l$ is assigned the interval with the maximum
  discrepancy out of the set $\{l_{left}, I_{left} \cup l_{right}\}$, and
  $r$ is assigned symmetrically to $l$.  The entire interval, $[0,1] =
  I:(i,l,r)$, is at the root of the tree, and the interval which maximizes
  the discrepancy is $i$.  Adding a point requires merging $O(\log n)$
  intervals if the tree is balanced, and the tree can be kept balanced
  through rotations which only require a constant number of merge
  operations each.
\end{proofsketch}

%\begin{lemma}
%Any general discrepancy function $\chi(p)$ can be maximized over rectangles in $O(n^2 \log n)$ time.
%\end{lemma}
%\begin{proof}
%  The algorithm of~\cite{DGM95} solves this problem in $O(n^2 \log n)$
%  time.  In the one-dimensional problem, each $I : (i, l, r)$ maintains
%  the same information, but under the measure $\Delta(\chi,\mathcal{R})$.
%  Since, $\Delta$ is additive, $l_{right} \cup r_{left}$ is still the
%  maximum discrepancy interval that spans the border of adjacent intervals
%  $I_{left}$ and $I_{right}$, thus the merge operation is correct.  This
%  implies that the dynamic one-dimensional algorithm is correct, and over
%  all slabs the maximum rectangle with maximum discrepancy is returned
%  correctly.
%\end{proof}

\begin{theorem}
\label{lemma:generallinear}
Let $\mathcal{R}^\prime$ be the set of all rectangles such that $\sum_{p
  \in R} m(p)$ and $\sum_{p \in R} b(p)$ are greater than the constant
$C$. Then any linear discrepancy function of the form $a_1 \sum m(p) + a_2
\sum b(p) + a_3$ can be maximized over this set in $O(n^2 \log n)$ time.
\end{theorem}
\begin{proof}
  Because $a_3$ is constant for all intervals, it can be ignored. Thus the
  linear function has the form of $d(m, b, R) = \sum_{p \in R} \chi(p)$.
  The algorithm of~\cite{DGM95} relies only on the fact that the
  discrepancy function is additive, and hence can be extended to the above
  discrepancy function by only modifying the intervals and merge operation
  in the one-dimensional case.  
  
  Define $\mathcal{I}^\prime$ to be the set
  of all intervals such that $\sum_{p \in I} m(p) \geq C$ and $\sum_{p \in
    I} b(p) \geq C$.  For each interval $I:(i,l,r)$, $i$ must be in
  $\mathcal{I}^\prime$ and $l$ and $r$ must represent sets of intervals
  $l_1 \ldots l_k$ and $r_1 \ldots r_h$, respectively.  $l_k$ (resp.
  $r_h$) is the interval in $\mathcal{I}^\prime$ which contains the left
  (resp. right) boundary that has the maximum discrepancy.  $l_1$ (resp.
  $r_1$) is the interval which contains the left (resp. right) boundary
  that has the maximum discrepancy.  For all $i$ $l_i$ includes the left
  boundary and $|l_i| < |l_j|$ for all $i<j$.  Also $\sum_{p \in l_i} m(p)
  < C$ or $\sum_{p \in l_i} b(p) < C$ for all $i<k$.
%, and if $\sum_{p \in l_i} m(p)\geq C$ (resp. $b(m)$), then $\sum_{p \in l_i} b(p) < \sum_{p \in l_{i+1}} b(p)$ (resp. $m(p)$).  
  Finally, the set $l$ must contain all local maximum; if $l_i$ were to
  gain on lose one point, the discrepancy would decrease.  The
  restrictions are the same for all $r_i$, expect these intervals must
  contain the right boundary.
  
  The local optimality restriction ensures $\sum_{p \in l_i} m(p) <
  \sum_{p \in l_{i+1}} m(p)$ and $\sum_{p \in l_i} b(p) < \sum_{p \in
    l_{i+1}} b(p)$.  If either measure ($\sum m(p)$ or $\sum b(p)$)
  increases then the other must also increase or this would violate the
  local optimality condition.  An increase of just a measure that
  increases discrepancy will cause the previous interval not to be optimal
  and an increase in just a measure that causes the discrepancy to
  decrease will cause the latter interval not to be optimal.  Thus $k$ and
  $h$ are constants bounded by the number of $p$ required for $\sum m(p)
  \geq C$ and $\sum b(p) \geq C$.  Thus each interval in the tree
  structure stores a constant amount of information.
  
  A merge between two adjacent intervals $I_{left}: (i_{left}, l_{left},
  r_{left})$ and $I_{right} : (i_{right}, l_{right}, r_{right})$ also can
  be done in constant time.  Computing the maximum discrepancy interval in
  $\mathcal{I}^\prime$ can be done by checking $i_{left}$ and $i_{right}$
  versus all pairs from $l_{right} \cup r_{left}$ such that $\sum m(p)
  \geq C$ and $\sum b(p) \geq C$.  There are a constant number of these.
  By the local optimality restriction, the optimal interval in
  $\mathcal{I}^\prime$ spanning the boundary must have one endpoint in
  each set.  Updating $l$ and $r$ can be done by just pruning from
  $l_{left}$ and $I_{left} \cup r_{left}$ according to the restrictions
  for $l$, and similarly for $r$.  These remain a constant size after the
  pruning.  Because a merge can be done in constant time, a point can be
  added to the balanced tree in $O(\log n)$ time.  Hence, the maximum
  discrepancy rectangle in $\mathcal{R}^\prime$ for any linear discrepancy
  function can be found in $O(n^2 \log n)$ time.
\end{proof}

A similar argument applies if we consider prospective discrepancy, or
higher dimensions. We omit details.
\begin{lemma}
A linear discrepancy function can be maximized over  
prospective rectangles in $O(n^3 \log n)$ time, or it can be maximized
over axis-parallel hyper-rectangles in $d$-dimensions in time
$O(n^{2d-2}\log n)$. 
\end{lemma}
%\begin{proof}
%  Trivial. For each point $p$, invoke Lemma~\ref{lemma:generallinear} on the set of
%  points with time greater than $t(p)$.
%\end{proof}

%%% Local Variables: 
%%% mode: latex
%%% TeX-master: "finalSODA"
%%% End: 

\section{One-parameter Exponential Families}
\label{sec:one-param-exp}
Having developed general algorithms for dealing with convex discrepancy
functions, we now present a general expression for a likelihood-based
discrepancy measure for the one-parameter exponential family. Many common
distributions like the Poisson, Bernoulli, Gaussian and gamma
distributions are members of this family. Subsequently we will derive
specific expressions for the above mentioned distribution families. 

\begin{defn}[One-parameter exp. family]
  The distribution of a random variable $y$ belongs to  a one-parameter
  exponential family (denoted by $y \sim \textrm{1EXP}(\eta,\phi,T,B_{e},a)$
  if it has probability density given by
\begin{equation*}
\label{oneparexp}
f(y;\eta)=C(y,\phi)exp((\eta T(y) - B_{e}(\eta))/a(\phi))
\end{equation*}
\noindent where $T(\cdot)$ is some measurable function, $a(\phi)$ is a function
of some known scale parameter $\phi(>0)$, $\eta$ is an unknown parameter (called the natural
parameter), and $B_{e}(\cdot)$ is a strictly convex function. The support
$\{y:f(y;\eta)>0\}$ is independent of $\eta$.
\end{defn}
\noindent It can be shown that $E_{\eta}(T(Y))=B_{e}^{'}(\eta)$ and
$\textrm{Var}_{\eta}(T(Y))=a(\phi)B_{e}^{''}(\eta)$. In general,  
$a(\phi) \propto \phi$. 

Let $\mathbf{y}=\{y_{i}:i \in R \}$ denote a set of $|R|$ variables that are
independently distributed with $y_{i} \sim
\textrm{1EXP}(\eta,\phi_{i},T,b,a)$, $(i \in R)$. The joint distribution of
$\mathbf{y}$ is given by
\begin{equation*}
\label{jointoneparexp} 
f(\mathbf{y};\eta)=\prod_{i \in R}C(y_{i},\phi_{i})exp((\eta T^{*}(\mathbf{y}) - B_{e}(\eta))/\phi^{*})
\end{equation*}
where $1/\phi^{*}=\sum_{i \in R}(1/a(\phi_{i}))$,
$v_i = \phi^* / a(\phi_i)$, and
%$v_{i}=a(\phi_{i})^{-1}/\sum_{j \in R}a(\phi_{j})^{-1}$, and
$T^{*}(\mathbf{y})=\sum_{i \in R}(v_{i}T(y_{i}))$.

Given data $\mathbf{y}$, the \emph{likelihood} of parameter $\eta$ is the
probability of seeing $\mathbf{y}$ if drawn from a distrbution with parameter
$\eta$. This function is commonly expressed in terms of its logarithm, the
\emph{log-likelihood} $l(\eta;\mathbf{y})$,  which is
given by (ignoring constants that do not depend on $\eta$)
\begin{equation}
\label{llik}
l(\eta;\mathbf{y})=(\eta T^{*}(\mathbf{y}) - B_{e}(\eta))/\phi^{*}
\end{equation}
and depends on data only through $T^{*}$ and $\phi^{*}$.

\begin{theorem}
\label{mle}
Let $\mathbf{y}=(y_{i}:i \in R)$ be independently distributed with
$y_{i} \sim \textrm{1EXP}(\eta,\phi_{i},T,b,a)$, $(i \in R)$. Then, the maximum
likelihood estimate (MLE) of $\eta$ is $\hat{\eta}=g_{e}(T^{*}(\mathbf{y}))$,
where $g_{e}=(B_{e}^{'})^{-1}$. The maximized log-likelihood (ignoring
additive constants) is
$l(\hat{\eta};\mathbf{y})=(T^{*}(\mathbf{y})g_{e}(T^{*}(\mathbf{y}))-B_{e}(g_{e}(T^{*}(\mathbf{y}))))/\phi^{*}$.
\end{theorem}

\begin{proof}
  The MLE is obtained by maximizing (\ref{llik}) as a function of $\eta$.
  Since $B_{e}$ is strictly convex, $B_{e}^{'}$ is strictly monotone and
  hence invertible. Thus, the MLE obtained as a solution of
  $l(\eta;\mathbf{y})^{'}=0$ is
  $\hat{\eta}=(B_{e}^{'})^{-1}(T^{*}(\mathbf{y}))=g_{e}(T^{*}(\mathbf{y}))$. The
  second part is obtained by substituting $\hat{\eta}$ in (\ref{llik}).
\end{proof}

The likelihood ratio test for outlier detection is based on the following
premise. Assume that data is drawn from a one-parameter exponential
family. For a given region $R_1$ and its complement $R_2$, let
$\eta_{R_1}$ and $\eta_{R_2}$ be the MLE parameters for the data in the
regions. Consider the two hypothesis $H_0: \eta_{R_{1}}=\eta_{R_{2}}$ and
$H_1:\eta_{R_{1}} \neq \eta_{R_{2}}$. The test then measures the ratio of
the likelihood of $H_1$ versus the likelihood of $H_0$. The resulting
quantity is a measure of the strength of $H_1$; the larger this number is,
the more likely it is that $H_1$ is true and that the region represents a
true outlier.  The likelihood ratio test is \emph{individually} the test
with most statistical power to detect the region of maximum discrepancy
and hence is optimal for the problems we consider. A proof of this fact
for Poisson distributions is provided by Kulldorff~\cite{kulldorff:comm} and
extends to $\textrm{1EXP}$ without modification.

%We are now ready to state the main theorem of this section.

\begin{theorem}
\label{llrt}
Let $\mathbf{y_{R_{j}}}=(y_{R_{j}i}:i \in R_{j})$ be independently
distributed with $y_{R_{j}i} \sim$
1EXP$(\eta_{R_{j}}$,$\phi_{R_{j}i}$,$T,B_{e},a)$, for $j=1,2$. The
log-likelihood ratio test statistic for testing
$H_{0}:\eta_{R_{1}}=\eta_{R_{2}}$ versus $H_{1}:\eta_{R_{1}} \neq
\eta_{R_{2}}$ is given by
\begin{equation}
\label{lrt}
\Delta = \kappa(G_{R_{1}},\Phi_{R_{1}})+\kappa(G_{R_{2}},\Phi_{R_{2}})-\kappa(G,\Phi)
\end{equation}
\noindent where $\kappa(x,y)=(x g_{e}(x) - B_{e}(g_{e}(x)))/y$, $G_{R_{j}}=T^{*}(\mathbf{y_{R_{j}}}), 
1/\Phi_{R_{j}}=\sum_{i \in R_{j}}(1/a(\phi_{R_{j}i}))$, 
$1/\Phi=1/\Phi_{R_{1}} + 1/\Phi_{R_{2}}$, $b_{R_{1}}=\frac{1/\Phi_{R_{1}}}{(1/\Phi_{R_{1}} + 1/\Phi_{R_{2}})}$ and 
$G=b_{R_{1}} G_{R_{1}} + (1-b_{R_{1}})G_{R_{2}}$.
\end{theorem}

\begin{proof}
The likelihood ratio is given by $\frac{sup_{\eta_{R_{1}} \neq
\eta_{R_{2}}}L(\eta_{R_{1}},\eta_{R_{2}};\mathbf{y_{R_{1}}},\mathbf{y_{R_{2}}})}{sup_{\eta}L(\eta;\mathbf{y_{R_{1}},
y_{R_{2}}})}$. Substituting the MLE expressions $\hat{\eta_{R_1}}$ and
$\hat{\eta_{R_2}}$ from Theorem~\ref{mle}, and setting 

% which equals
%$\frac{L(\hat{\eta_{R_{1}}};\mathbf{y_{R_{1}}})L(\hat{\eta_{R_{2}}};\mathbf{y_{R_{2}}})}{L(\hat{\eta};\mathbf{y_{R_{1}},y_{R_{2}}})}$
%with the MLEs obtained from Theorem~\ref{mle}. Also,
$G=T^{*}(\mathbf{y_{R_{1}},y_{R_{2}}})=\frac{\sum_{j=1,2}\sum_{i \in
R_{j}}T(y_{R_{j}i})/a(\phi_{R_{j}i})}{\sum_{j=1,2}\sum_{i \in
R_{j}}1/a(\phi_{R_{j}i})} =\frac{1/\Phi_{R_{1}}}{(1/\Phi_{R_{1}} +
1/\Phi_{R_{2}})}G_{R_{1}} + \frac{1/\Phi_{R_{2}}}{(1/\Phi_{R_{1}} +
1/\Phi_{R_{2}})}G_{R_{2}} = b_{R_{1}} G_{R_{1}} + (1-b_{R_{1}})G_{R_{2}}$,
the result follows by computing logarithms.% and substituting the maximum
%likelihood expressions from Theorem~\ref{mle}.
\end{proof}

\begin{fact}
To test $H_{0}:\eta_{R_{1}}=\eta_{R_{2}}$ versus $H_{1}:\eta_{R_{1}} >
\eta_{R_{2}}$, the log-likelihood ratio test statistic is given by
\begin{equation}
\label{onesidedlrt}
\Delta = 1(\hat{\eta_{R_{1}}} > \hat{\eta_{R_{2}}})(\kappa(G_{R_{1}},\Phi_{R_{1}})+\kappa(G_{R_{2}},\Phi_{R_{2}})-\kappa(G,\Phi))
\end{equation}
Similar result holds for the alternative $H_{1}:\eta_{R_{1}} < \eta_{R_{2}}$ with the inequalities reversed.
\end{fact}

In the above expression for $\Delta$ (with $R_1=R,R_2=R^c$), the key
terms are the values $b_R$ and $G_R$. $G_R = T^*(\mathbf{y_R})$ is a
function of the data ($T^*$ is a
\emph{sufficient statistic} for the distribution), and thus is the
equivalent of a measurement. In fact, $G_R$ is a weighted average
of  $T(y_{i})$s in $R$. Thus, $G_R/\Phi_R=\sum_{i \in R}T(y_{i})/a(\phi_{i})$
represents the \emph{total} in $R$. Similarly, $G/\Phi$ gives
the aggregate for the region and hence $m_R=\frac{\Phi}{\Phi_R}\frac{G_R}{G}$
is the fraction of total contained in $R$. Also, $1/\Phi_{R}$ gives
the total area of $R$ which is independent of the actual measurements and
only depends on some baseline measure. Hence, $b_R=\frac{\Phi}{\Phi_R}$ gives
the fraction of total area in R. The next theorem provides an
expression for $\Delta$ in terms of $m_R$ and $b_R$. 

%The term $b_R$ represents distributional
%parameters, and corresponds to baseline values for the data.  
%To express
%$\Delta$ in terms of $b_R$ and true measurement values $m_R$, we
%reparametrize $\Delta$, substituting $m_R  = b_R G_R/G$ for $G_R$. 

\begin{theorem}
\label{alt-param}
Let $R_{1}=R$ and $R_{2}=R^{c}$. To obtain $R \in {\cal R}$ that maximizes
discrepancy, assume $G$ and $\Phi$ to be fixed and consider the
parametrization of $\Delta$ in terms of $b_{R}$ and $m_{R}=b_{R} G_{R}/G$.

The discrepancy measure (ignoring additive constants) $d(.,.)$ is given by
\begin{eqnarray}
\nonumber 
d(m_{R},b_{R})\frac{\Phi}{G} 
&=& 
m_{R} g_{e}(G\frac{m_{R}}{b_{R}})-\frac{b_{R}}{G}B_{e}(g_{e}(G\frac{m_{R}}{b_{R}})) +
\\ \label{repar} & &
(1-m_{R}) g_{e}(G\frac{1-m_{R}}{1-b_{R}})- 
\\ \nonumber & &
\frac{(1-b_{R})}{G}B_{e}(g_{e}(G\frac{1-m_{R}}{1-b_{R}})) 
\end{eqnarray}
\end{theorem}

\begin{proof}
Follows by substituting $G_{R}=G\frac{m_{R}}{b_{R}}$,
$G_{R^{c}}=G\frac{1-m_{R}}{1-b_{R}}$ in (\ref{lrt}), simplifying 
and ignoring additive constants.
\end{proof}

\section{Discrepancy Measures For Specific Distributions}

We can now put together all the results from the previous
sections. Section~\ref{sec:conv-appr-theor} showed how to map a convex
discrepancy function to a collection of linear discrepancy functions, and
Section~\ref{sec:algos} presented algorithms maximizing general linear
discrepancy functions over axis parallel rectangles. The previous section
presented a general formula for discrepancy in a one-parameter exponential
family. We will now use all these results to derive discrepancy functions
for specific distribution families and compute maximum discrepancy
rectangles with respect to them.

\subsection{The Kulldorff Scan Statistic (Poisson distribution)}
\label{ssec:poisson-deriv}

The Kulldorff scan statistic was designed for data generated by an
underlying Poisson distribution. We reproduce Kulldorff's derivation of
the likelihood ratio test, starting from our general discrepancy function
$\Delta$. 

In a Poisson distribution, underlying points are marked for the presence of
some rare event (e.g. presence of some rare disease in an individual) and
hence the measurement attached to each point is binary with a $1$ indicating
presence of the event. The number of points that get marked on a region
$R$ follows a Poisson process with base measure $b$ and intensity
$\lambda$ if (i) $N(\emptyset)=0$, (ii)  $N(A) \sim \textrm{Poisson}(\lambda
b(A)), A \subset R, b(\cdot)$ is a baseline measure defined on $R$ and
$\lambda$ is a fixed intensity parameter (examples of $b(A)$ include the
area of $A$, total number of points in $A$, \emph{etc.}), and (iii) the number
of marked points in disjoint subsets are independently distributed.

\paragraph{Derivation of the Discrepancy Function.}
A random variable $y \sim \textrm{Poisson}(\lambda\mu)$ is a member of
$\textrm{1EXP}$ with $T(y)=y/\mu, \phi=1/\mu, a(\phi)=\phi,
\eta=\log(\lambda), B_{e}(\eta)=\exp(\eta), g_{e}(x)=\log(x)$. For a set
of $n$ independent measurements with mean $\lambda\mu_{i},i=1,\cdots,n$,
$T^{*}(\mathbf{y})=\sum_{i=1}^{n}y_{i}/\sum_{i=1}^{n}\mu_{i},
\phi^{*}=(\sum_{i=1}^{n}\mu_{i})^{-1}$.  For a subset $R$, assume the
number of marked points follows a Poisson process with base measure
$b(\cdot)$ and log-intensity $\eta_{R}$ while that in $R^{c}$ has the same
base measure but log-intensity $\eta_{R^{c}}$.  For any partition
$\{A_{i}\}$ of $R$ and $\{B_{j}\}$ of $R^{c}$, $\{N(A_{i})\}$ and
$\{N(B_{j})\}$ are independently distributed Poisson variables with mean
$\{\exp(\eta_{R})b(A_{i})\}$ and $\{\exp(\eta_{R^{c}})b(B_{j})\}$
respectively. Then,
$1/\Phi_{R}=\sum_{A_{i}}b(A_{i}))=b(R)$, $1/\Phi_{R^{c}}=b(R^{c})$, 
$G_{R}=\frac{\sum_{A_{i}}N(A_{i})}{\sum_{A_{i}}b(A_{i})}=N(R)/b(R)$,
$G_{R^{c}}=N(R^{c})/b(R^{c})$,
and $G=\frac{N(R) + N(R^{c})}{b(R) + b(R^{c})}$.  Hence,
$b_{R}=\frac{b(R)}{b(R) +  b(R^{c})}$ and $m_{R}=\frac{N(R)}{N(R) + N(R^{c})}$. 
\begin{eqnarray*}
d_K(b_{R},m_{R})\frac{\Phi}{G}
&=&
m_{R}(\log(G) + \log(\frac{m_{R}}{b_{R}})) - b_{R}\frac{m_{R}}{b_{R}} +
\\ \nonumber &&
(1-m_{R})(\log(G) + \log(\frac{1-m_{R}}{1-b_{R}}))-
\\ \nonumber && 
\frac{1-m_{R}}{1-b_{R}}(1-b_{R})
\\ &=& m_{R} \log(\frac{m_{R}}{b_{R}}) + 
\\ &&
(1-m_{R})\log(\frac{1-m_{R}}{1-b_{R}}) + const 
\end{eqnarray*}
and hence $d_K(b_{R},m_{R}) = c(m_{R} \log(\frac{m_{R}}{b_{R}}) +
(1-m_{R})\log(\frac{1-m_{R}}{1-b_{R}}))$, where $c >0$ is a fixed
constant. Note that the discrepancy is independent of the partition used
and hence is well defined.

\paragraph{Maximizing the Kulldorff Scan Statistic.}

It is easy to see that $d_K$ is a convex function of $m_R$ and $b_R$, is
always positive, and grows without bound as either of $m_R$ and $b_R$
tends to zero. It is zero when $m_R = b_R$. The Kulldorff scan statistic
can also be viewed as the Kullback-Leibler distance between the two
two-point distributions $[m_R, 1-m_R]$ and $[b_R, 1-b_R]$. 
As usual, we will consider maximizing the Kulldorff scan statistic over
the region $S_n = [1/n, 1-1/n]^2$. To estimate the size of an
$\epsilon$-approximate family for $d_K$, we will compute $\lambda^*$ over
$S_n$. 

Let $f_K(x,y) = x \ln \frac{x}{y} + (1-x)\ln \frac{1-x}{1-y}$. 
\begin{eqnarray*}
 \nabla f_K &=& \mathbf{i}\left( \ln \frac{x}{1-x}  + \ln \frac{y}{1-y} \right) +  
% \\ \nonumber && 
 \mathbf{j}\left( \frac{x}{y} - \frac{1-x}{1-y} \right)
\end{eqnarray*}

\begin{align*}
  H(f_K) =
  \begin{pmatrix}
  \frac{\partial^2 f}{\partial x^2} & \frac{\partial^2 f}{\partial x \partial y} \\
  \frac{\partial^2 f}{\partial y \partial x} & \frac{\partial^2 f}{\partial y^2}
  \end{pmatrix} = 
  \begin{pmatrix}
    \frac{1}{x(1-x)} & \frac{1}{y(1-y)}  \\
    \frac{1}{y(1-y)} & \frac{-x}{y^2} - \frac{1-x}{(1-y)^2} \\
  \end{pmatrix}
\end{align*}

The eigenvalues of $H(f)$ are the roots of the equation $| H(f) - \lambda
\mathbf{I} | = 0$. Solving for $\lambda^*$, and substituting from the
expressions for the partial derivatives, and maximizing over $S_n$, we
obtain $\lambda^* = \Theta(n)$.

Invoking Theorem~\ref{thm:main-approx} and Theorem~\ref{lemma:generallinear},

\begin{theorem}
\label{thm:kulldorff-alg}
  An additive $\epsilon$-approximation to the maximum discrepancy $d_K$
  over all rectangles containing at 
  least a constant measure can be computed in time
  $O(\frac{1}{\epsilon}n^2 \log^2   n)$. With respect to prospective time
  windows, the corresponding maximization takes time
  $O(\frac{1}{\epsilon}n^3\log^2 n)$. 
\end{theorem}

The Jensen-Shannon divergence is a symmetrized variant of the
Kullback-Leibler distance. We mentioned earlier that $d_K$ can be
expressed as a Kullback-Leibler distance. Replacing this by the
Jensen-Shannon distance, we get a symmetric version of the Kulldorff
statistic, for which all the bounds of Theorem~\ref{thm:kulldorff-alg}
apply directly.

\subsection{Gaussian Scan Statistic}

It is more natural to use an underlying Gaussian process when measurements
are real numbers, instead of binary events. In
this section, we derive a discrepancy function for an underlying Gaussian
process. To the best of our knowledge, this derivation is novel.

\paragraph{Derivation of the Discrepancy Function.} 
A random variable $y$ that follows a Gaussian distribution with mean $\mu$
and variance $1/\tau^{2}$ (denoted as $y \sim N(\mu,1/\tau^{2})$ is a
member of $\textrm{1EXP}$ with $T(y)=y, \eta=\mu, B_{e}(\eta)=\eta^2/2,
\phi=1/\tau^{2}, a(\phi)=\phi, g_{e}(x)=x$. For a set of $n$ independent
measurements with mean $\mu$ and variances
$1/\tau^{2}_{i},i=1,\cdots,n$(known),
$\phi^{*}=(\sum_{i=1}^{n}\tau^{2}_{i})^{-1}$ and
$T^{*}(\mathbf{y})=\sum_{i=1}^{n}y_{i}\tau_{i}^{2}/\sum_{i=1}^{n}\tau_{i}^{2}$.
Assume measurements in $R$ are independent $N(\mu_{R},1/\tau_{i}^{2}),(i
\in R)$ while those in $R^{c}$ are independent
$N(\mu_{R^{c}},1/\tau_{i}^{2}),(i \in R^{c})$. Then, $\Phi_{R}=(\sum_{i
  \in R}\tau_{i}^{2})^{-1}$,$\Phi_{R^{c}}=(\sum_{i \in
  R^{c}}\tau_{i}^{2})^{-1}$, $G_{R}=\frac{\sum_{i \in
    R}\tau_{i}^{2}y_{i}}{\sum_{i \in R}\tau_{i}^{2}}$,
$G_{R^{c}}=\frac{\sum_{i \in R^{c}}\tau_{i}^{2}y_{i}}{\sum_{i \in
    R^{c}}\tau_{i}^{2}}$, and $G=\frac{\sum_{i \in
    R+R^{c}}\tau_{i}^{2}y_{i}}{\sum_{i \in R+R^{c}}\tau_{i}^{2}}$. 
Hence,
$b_{R} = \frac{1/\Phi_{R}}{(1/\Phi_{R} + 1/\Phi_{R^{c}})} = \frac{\sum_{i
    \in R}\tau_{i}^{2}}{\sum_{i \in R+R^{c}}\tau_{i}^{2}}$ and
$m_{R}=\frac{\sum_{i \in R}\tau_{i}^{2}y_{i}}{\sum_{i \in
    R+R^{c}}\tau_{i}^{2}}$. Thus,

\begin{eqnarray*}
\lefteqn{d_G(b_{R},m_{R})\frac{\Phi}{G} = m_{R} G \frac{m_{R}}{b_{R}} - \frac{b_{R}}{G}G\frac{m_{R}}{b_{R}} +}
\\ && 
(1-m_{R})G\frac{1-m_{R}}{1-b_{R}}- \frac{1-b_{R}}{G}G\frac{1-m_{R}}{1-b_{R}} 
\\ &=&
G(\frac{m_{R}^{2}}{b_{R}} + \frac{(1-m_{R})^{2}}{1-b_{R}}) - 1
=
G\frac{(m_{R}-b_{R})^{2}}{b_{R}(1-b_{R})} 
\end{eqnarray*}
and hence $d_G(b_{R},m_{R}) = c\frac{(m_{R}-b_{R})^{2}}{b_{R}(1-b_{R})}$,
  where $c>0$ is a fixed constant. %An important special case occurs when
%  the variance is constant (i.e. $\tau_{i}^{2}=\tau^{2}$ for each $i$).
%  Then, $b_{R}=|R|/(|R| + |R^{c}|)$ and $m_{R} = \frac{\sum_{i \in
%      R}y_{i}}{\sum_{i \in R+R^{c}}y_{i}}$ with the discrepancy being
%  maximized if a small fraction of points account for a large fraction of
%  the total or vice-versa.  
Note that the underlying baseline $b(\cdot)$
  is a weighted counting measure which aggregate weights $\tau_{i}^{2}$
  attached to points in a region.

\paragraph{Maximizing the Gaussian Scan Statistic.}

Again, it can be shown that $d_G$ is a convex function of both parameters,
and grows without bound as $b_R$ tends to zero or one. Note that 
this expression can be viewed as the $\chi^2$-distance between the two
two-point distributions $[m_R, 1-m_R], [b_R, 1-b_R]$. 
The complexity of an $\epsilon$-approximate family for $d_G$ can be
analyzed as in Section~\ref{ssec:poisson-deriv}. Let $f_G(x,y) =
\frac{(x-y)^2}{y(1-y)}$.
% \begin{align*}
%  \nabla f_G &= \mathbf{i}\left(\frac{2x-2y}{y(1-y)}\right) +
%  \mathbf{j}\left( \frac{2y-2x}{y(1-y)} -
%  \frac{(x-y)^2 (1-2y)}{y^2 (1-y)^2} \right) 
%\end{align*}
%\begin{eqnarray*}
%\lefteqn{H(f_G) = } \\&&
%\begin{pmatrix}
%    \frac{2}{y(1-y)} & \frac{-2}{y(1-y)} - \frac{2(x-y)(1-2y)}{y^2(1-y)^2} \\
%    \frac{-2}{y(1-y)} - \frac{2(x-y)(1-2y)}{y^2(1-y)^2} & 
%    \frac{2}{y(1-y)} + \frac{4(x-y)(1-2y) - 2(x-y)^2}{y^2(1-y)^2} \\
%    & + \frac{(x-y)^2(1-2y)^2}{y^3(1-y)^3}\\
%\end{pmatrix}
%\end{eqnarray*}
Expressions for $\nabla f_G$ and $H(f_G)$ are presented in Appendix~\ref{ssec:g}.
Solving the equation $| H - \lambda \mathbf{I}|$, and maximizing
over $S_n$, we get $\lambda^* = O(n^2)$.

\begin{theorem}
  An additive $\epsilon$-approximation to the maximum discrepancy $d_G$
  over all rectangles containing at 
  least a constant measure can be computed in time
  $O(\frac{1}{\epsilon}n^3 \log n\log\log n)$. With respect to prospective time
  windows, the corresponding maximization takes time
  $O(\frac{1}{\epsilon}n^4 \log n\log\log n)$. 
\end{theorem}

\paragraph{Trading Error for Speed}
\label{ssec:relative-error}

For the Kulldorff statistic, the function value grows slowly as it
approaches the boundaries of $S_n$. Thus, only minor improvements can be
made when considering relative error approximations. However, for the
Gaussian scan statistic, one can do better. A simple substitution shows
that when $x = 1 - \frac{1}{n}$, $y = \frac{1}{n}$, $f_G(x,y) =
\Theta(n)$. Using this bound in Theorem~\ref{thm:main-approx}, we see that
a relative $\epsilon$-approximate family of size
$O(\frac{1}{\epsilon}\log n)$ can be constructed for $d_G$, thus
yielding the following result:

\begin{theorem}
  A $1/(1+\epsilon)$ approximation to the maximum discrepancy $d_G$ over the
  space of axis parallel rectangles containing constant measure can be
  computed in time $O(\frac{1}{\epsilon}n^2\log^2 n)$.
\end{theorem}

\subsection{Bernoulli Scan Statistic}

Modeling a system with an underlying Bernoulli distribution is appropriate when the events are binary, but more common than those that would be modeled with a Poisson distribution.  For instance, a baseball player's batting average may describe a Bernoulli distribution of the expectation of a hit, assuming each at-bat is independent.  

%Similar derivations can be made for other well known distributions. We
%state the results without proofs.
%\begin{description}

%\item[Bernoulli Distribution] \begin{eqnarray*}
%d_B(b_{R},m_{R})\frac{\Phi}{G}&=&m_{R} \log(\frac{m_{R}}{b_{R}}) + (1-m_{R})\log(\frac{1-m_{R}}{1-b_{R}})\\
%&+& (\frac{b_{R}}{G}-m_{R})\log(1-G\frac{m_{R}}{b_{R}}) + (\frac{1-b_{R}}{G} -1 + m_{R})\log(1 - G\frac{1-m_{R}}{1-b_{R}}) \\
%\end{eqnarray*}

%\item[Gamma distribution] \begin{eqnarray*}
%d_\gamma(b_{R},m_{R})\frac{\Phi}{G}&=&m_{R}(-\frac{b_{R}}{Gm_{R}})-\frac{b_{R}}{G}\log(G\frac{m_{R}}{b_{R}})
%+(1-m_{R})(-\frac{1-b_{R}}{G(1-m_{R})})-\frac{1-b_{R}}{G}\log(G\frac{1-m_{R}}{1-b_{R}})\\
%%&=&b_{R} \log(\frac{b_{R}(1-m_{R})}{m_{R}(1-b_{R})})-\log(\frac{1-m_{R}}{1-b_{R}}) + const\\
%&=&b_{R} \log(\frac{b_{R}}{m_{R}}) + (1-b_{R})\log(\frac{1-b_{R}}{1-m_{R}}) + const
%\end{eqnarray*} 
%and hence ignoring additive constants, $d_\gamma(b_{R},m_{R})=c(b_{R} \log(\frac{b_{R}}{m_{R}}) + (1-b_{R})\log(\frac{1-b_{R}}{1-m_{R}})),
%c(>0)$ is a fixed constant.
%\end{description}

\paragraph{Derivation of the Discrepancy Function.}
 A
binary measurment $y$ at a point has a Bernoulli distribution with
parameter $\theta$ if $P(y=1)=\theta^{y}(1-\theta)^{1-y}$.  This is a
member of $\textrm{1EXP}$ with
$T(y)=y,\eta=\log(\frac{\theta}{1-\theta}),B_{e}(\eta)=\log(1+\exp(\eta)),\phi=1,a(\phi)=1,
g_{e}(x)=\log(x)-\log(1-x)$.

For a set of $n$ independent measurements with parameter $\eta$, $\phi^{*}=1/n, T^{*}(\mathbf{y})=\sum_{i=1}^{n}y_{i}/n$. 
Assuming measurements in $R$ and $R^{c}$ are independent Bernoulli with parameters $\eta_{R}$ and $\eta_{R^{c}}$ respectively, $\Phi_{R}=1/|R|,\Phi_{R^{c}}=1/|R^{c}|, G_{R}=y(R)/|R|, G_{R^{c}}=y(R^{c})/|R^{c}|, b_{R}=\frac{|R|}{|R|+|R^{c}|}, G=\frac{y(R) + y(R^{c})}{|R|+|R^{c}|}, m_{R}=\frac{y(R)}{y(R) + y(R^{c})}$. 
Note that $y(A)$ denotes the number of 1's in a subset $A$. Thus,
\begin{eqnarray*}
\lefteqn{d_B(b_{R},m_{R})\frac{\Phi}{G} = m_{R} \log(\frac{m_{R}}{b_{R}}) +}
\\ &&
 (1-m_{R})\log(\frac{1-m_{R}}{1-b_{R}}) +
%\\ && 
(\frac{b_{R}}{G}-m_{R})\log(1-G\frac{m_{R}}{b_{R}})
\\ &&
+ (\frac{1-b_{R}}{G} -1 + m_{R})\log(1 - G\frac{1-m_{R}}{1-b_{R}})
\end{eqnarray*}

\paragraph{Maximizing the Bernoulli Scan Statistic.}
Much like $d_K$, it is easy to see that $d_B$ is a convex function of $m_R$ and $b_R$, is always positive, and grows without bound as either $b_R$ or $m_R$ tend to zero or one.  
The complexity of an $\epsilon$-approximate family for $d_B$, the
Bernoulli scan statistic, can be analyzed by letting $f_B(x,y) = x \log \frac{x}{y} + (1 - x) \log \frac{1-x}{1-y} + \left(\frac{y}{G} - x\right) \log \left(1 - G\frac{x}{y}\right) + \left(\frac{1-y}{G} - 1 + x\right) \log \left(1 - G \frac{1-x}{1-y}\right)$, where $G$ is a constant. 
%\begin{figure*}[htbp]
%\label{fig:bernoulli}
%\begin{align*}
%  \nabla f_B =& \mathbf{i}
% \left(\begin{array}{l}
% \log \frac{x}{y} - \log \frac{1-x}{1-y} + \log \left(1 - G \frac{1-x}{1-y}\right) - \\
% \log \left(1 - G\frac{x}{y}\right)
% \end{array}\right) +
% \\ &
%  \mathbf{j} \left(\frac{1}{G} \log \left(1 - G \frac{x}{y}\right) - \frac{1}{G} \log \left(1 - G\frac{1-x}{1-y} \right)
%  \right) 
%\end{align*}
%\begin{gather*}
%  H(f_B) =
%  \begin{pmatrix}
%    \frac{1}{x} + \frac{1}{1-x} + \frac{G}{(1-y) - G(1-x)} + \frac{G}{y -
%      G x} &
%    \frac{-1}{y - Gx} - \frac{1}{(1-y) - G(1-x)} \\
%    \frac{-1}{y - Gx} - \frac{1}{(1-y) - G(1-x)} & \frac{-1}{y} +
%    \frac{1}{1-y} - \frac{G(1-x)}{(1-y)((1-y) - G(1-x))} -
%    \frac{Gx}{y(y-Gx)}
%  \end{pmatrix}
%\end{gather*}
%\caption{Expressions for $\nabla f_B$ and $H(f_B)$}
%\end{figure*}
The expressions for $\nabla f_B$ and $H(f_B)$ are presented in
Appendix~\ref{ssec:b}.  Direct substitution of the parameters yields $\lambda^* = O(n)$. 

\begin{theorem}
  An additive $\epsilon$-approximation to the maximum discrepancy $d_B$
  over all rectangles containing at 
  least a constant measure can be computed in time
  $O(\frac{1}{\epsilon}n^2 \log^2 n)$. With respect to prospective time
  windows, the corresponding maximization takes time
  $O(\frac{1}{\epsilon}n^3 \log^2 n)$. 
\end{theorem}

\subsection{Gamma Scan Statistic}

When events arrive one after another, where a Poisson variable describes the interval between events, then a gamma distribution describes the count of events after a set time.  

\paragraph{Derivation of the Discrepancy Function.}

A positive measurement $y$ has a gamma distribution with mean $\mu(>0)$ and shape $\nu(>0)$ if it has density 
$\frac{\nu^{\nu}}{\mu^{\nu}\Gamma(\nu)}\exp(-\frac{\nu}{\mu}y)x^{\nu-1}$ 
and is a member of $\textrm{1EXP}$ with 
$T(y) = y, \eta=-\frac{1}{\mu}(<0),B_{e}(\eta)=-\log(-\eta),\phi=1/\nu,a(\phi)=\phi,g_{e}(x)=-\frac{1}{x}$. Following arguments similar to the Gaussian case, 
$\Phi_{R}=(\sum_{i \in R}\nu_{i})^{-1}, 
\Phi_{R^{c}}=(\sum_{i \in R^{c}}\nu_{i})^{-1}, 
G_{R}=\frac{\sum_{i \in R}\nu_{i}y_{i}}{\sum_{i \in R}\nu_{i}}, 
G_{R^{c}}=\frac{\sum_{i \in R^{c}}\nu_{i}y_{i}}{\sum_{i \in R^{c}}\nu_{i}}, 
G=\frac{\sum_{i \in R+R^{c}}\nu_{i}y_{i}}{\sum_{i \in R+R^{c}}\nu_{i}}$. 
Hence, 
$b_{R} = \frac{1/\Phi_{R}}{(1/\Phi_{R} + 1/\Phi_{R^{c}})} 
= \frac{\sum_{i \in R}\nu_{i}}{\sum_{i \in R+R^{c}}\nu_{i}}$ 
and 
$m_{R}=\frac{\sum_{i \in R}\nu_{i}y_{i}}{\sum_{i \in R+R^{c}}\nu_{i}y_{i}}$. 
Thus,
\begin{eqnarray*}
\lefteqn{d_\gamma(b_{R},m_{R})\frac{\Phi}{G} =
m_{R}(-\frac{b_{R}}{Gm_{R}})-\frac{b_{R}}{G}\log(G\frac{m_{R}}{b_{R}}) +}
\\ &&
(1-m_{R})(-\frac{1-b_{R}}{G(1-m_{R})})-\frac{1-b_{R}}{G}\log(G\frac{1-m_{R}}{1-b_{R}})
\\ &=&
b_{R} \log(\frac{b_{R}(1-m_{R})}{m_{R}(1-b_{R})})-\log(\frac{1-m_{R}}{1-b_{R}}) + const\\
\\ &=&
b_{R} \log(\frac{b_{R}}{m_{R}}) + (1-b_{R})\log(\frac{1-b_{R}}{1-m_{R}}) + const
\end{eqnarray*} 
and hence ignoring additive constants, $d_\gamma(b_{R},m_{R})=c(b_{R} \log(\frac{b_{R}}{m_{R}}) + (1-b_{R})\log(\frac{1-b_{R}}{1-m_{R}})),
c(>0)$ is a fixed constant. For a fixed shape parameter (i.e. $\nu_{i}=\nu$ for each $i$),
$b_{R} =\frac{|R|}{|R|+|R^{c}|}$ and $m_{R}=\frac{\sum_{i \in R}y_{i}}{\sum_{i \in R+R^{c}}y_{i}}$.

\paragraph{Maximizing the Gamma Scan Statistic.}

Because $d_\gamma = d_K$ up to an additive constant, $f_\gamma = f_K$ and thus $\lambda^* = O(n)$ for $H(f_\gamma)$.  

\begin{theorem}
  An additive $\epsilon$-approximation to the maximum discrepancy $d_\gamma$
  over all rectangles containing at 
  least a constant measure can be computed in time
  $O(\frac{1}{\epsilon}n^2 \log^2 n)$. With respect to prospective time
  windows, the corresponding maximization takes time
  $O(\frac{1}{\epsilon}n^3 \log^2 n)$. 
\end{theorem}

%Both these discrepancy functions have Hessians whose largest eigenvalue
%over $S_n$ grows as $O(n)$ (see Sections~\ref{ssec:b},~\ref{ssec:gamma}). Thus, all the results from
%Section~\ref{ssec:poisson-deriv} apply identically in these cases as
%well. 
%\input gen-streaming-lowerbound

\bibliographystyle{acm}
\bibliography{arxiv1}

\appendix

\section{Gradients and Hessians}
\label{sec:eig}

%\subsection{The Kulldorff Scan Statistic}
%\label{ssec:k}

%Let $f_K(x,y) = x \ln \frac{x}{y} + (1-x)\ln \frac{1-x}{1-y}$. 
%$  \nabla f_K = \mathbf{i}( \ln x - \ln (1-x) + \ln y - \ln (1-y)) +  \mathbf{j}( \frac{x}{y} - \frac{1-x}{1-y}  )$.

%\begin{align*}
%  H(f_K) = 
%  \begin{pmatrix}
%    \frac{\partial^2 f}{\partial x^2} & \frac{\partial^2 f}{\partial x\partial y} \\
%\frac{\partial^2 f}{\partial y\partial x} & \frac{\partial^2 f}{\partial y^2} \\
%  \end{pmatrix} = 
%  \begin{pmatrix}
%    \frac{1}{x(1-x)} & \frac{1}{y(1-y)}  \\
%    \frac{1}{y(1-y)} & \frac{-x}{y^2} - \frac{1-x}{(1-y)^2} \\
%  \end{pmatrix}
%\end{align*}

%The eigenvalues of $H(f)$ are the roots of the equation $| H(f) - \lambda
%\mathbf{I} | = 0$. Solving for $\lambda^*$, and substituting from the
%expressions for the partial derivatives, and maximizing over $S_n$, we
%obtain $\lambda^* = \Theta(n)$.

\subsection{Gaussian Scan Statistic}
\label{ssec:g}

Recall that 
$\displaystyle{f_G(x,y) = \frac{(x-y)^2}{y(1-y)}}$. 
 \begin{align*}
   \nabla f_G &= \mathbf{i}\left(\frac{2x-2y}{y(1-y)}\right) +
   \mathbf{j}\left( \frac{2y-2x}{y(1-y)} -
   \frac{(x-y)^2 (1-2y)}{y^2 (1-y)^2} \right) 
 \end{align*}
 \begin{align*}
   H(f_G) = 
   \begin{pmatrix}
     \frac{2}{y(1-y)} & 
     \frac{-2}{y(1-y)} - \frac{2(x-y)(1-2y)}{y^2(1-y)^2} \\
     \frac{-2}{y(1-y)} - \frac{2(x-y)(1-2y)}{y^2(1-y)^2} &
     \frac{2}{y(1-y)} + \frac{4(x-y)(1-2y) - 2(x-y)^2}{y^2(1-y)^2} + \frac{(x-y)^2(1-2y)^2}{y^3(1-y)^3} 
   \end{pmatrix}
 \end{align*}
%Solving the equation $| H - \lambda \mathbf{I}|$ as above, and maximizing over $S_n$, we get $\lambda^* = O(n^2)$. 

\subsection{Bernoulli Scan Statistic}
\label{ssec:b}
Recall that 
$$f(x,y) = x \log \frac{x}{y} + (1 - x) \log \frac{1-x}{1-y} + \left(\frac{y}{G} - x\right) \log \left(1 - G\frac{x}{y}\right) + \left(\frac{1-y}{G} - 1 + x\right) \log \left(1 - G \frac{1-x}{1-y}\right),$$
where $G$ is a constant. 

\begin{align*}
  \nabla f =& \mathbf{i}\left(\log \frac{x}{y} - \log \frac{1-x}{1-y} + \log \left(1 - G \frac{1-x}{1-y}\right) - \log \left(1 - G\frac{x}{y}\right)\right) +\\ & \mathbf{j} \left(\frac{1}{G} \log \left(1 - G \frac{x}{y}\right) - \frac{1}{G} \log \left(1 - G\frac{1-x}{1-y} \right)
  \right) 
\end{align*}

\begin{align*}
  H(f) = 
  \begin{pmatrix}
    \frac{1}{x} + \frac{1}{1-x} + \frac{G}{(1-y) - G(1-x)} + \frac{G}{y - G x} & 
    \frac{-1}{y - Gx} - \frac{1}{(1-y) - G(1-x)} \\
    \frac{-1}{y - Gx} - \frac{1}{(1-y) - G(1-x)} & 
    \frac{-1}{y} + \frac{1}{1-y} - \frac{G(1-x)}{(1-y)((1-y) - G(1-x))} - \frac{Gx}{y(y-Gx)} 
  \end{pmatrix}
\end{align*}

%Direct substitution of the parameters yields $\lambda^* = O(n)$. 

%\subsection{Gamma Scan Statistic}
%\label{ssec:gamma}

%The gamma scan statistic has the same form as $d_K$, and the same analysis
%applies as in Section~\ref{ssec:k}. 

\end{document}